\documentclass[conference]{IEEEtran}
\IEEEoverridecommandlockouts
\usepackage{cite}
\usepackage{amsmath,amssymb,amsfonts}
\usepackage{algorithm}
\usepackage{algorithmic}
\usepackage{graphicx}
\usepackage{textcomp}
\usepackage{enumitem}
\usepackage{graphicx}
\usepackage{hyperref}
\usepackage{caption}
\usepackage{booktabs}
\usepackage{tabularx}
\usepackage{diagbox} 
\usepackage[most]{tcolorbox}

\usepackage[table]{xcolor}
\usepackage{array}
\usepackage{makecell}

\usepackage{diagbox}
\usepackage{multirow}
\usepackage{colortbl} 
\definecolor{Gray}{gray}{0.92}
\definecolor{LightCyan}{rgb}{0.88,1,1}
\arrayrulecolor{black}
\usepackage{hhline}

\usepackage{multirow}
\usepackage[flushleft]{threeparttable}
\usepackage{tablefootnote}
\renewcommand\IEEEkeywordsname{Index Terms}
\def\BibTeX{{\rm B\kern-.05em{\sc i\kern-.025em b}\kern-.08em
    T\kern-.1667em\lower.7ex\hbox{E}\kern-.125emX}}

\graphicspath{{Images/}}

\begin{document}


\title{DTM: Deterministic Approaches for Black-box Test Suite Minimization with Tree-based Similarity}

\makeatletter
\newcommand{\linebreakand}{%
  \end{@IEEEauthorhalign}
  \hfill\mbox{}\par
  \mbox{}\hfill\begin{@IEEEauthorhalign}
}
\makeatother

\author{\IEEEauthorblockN{Md. Siam}
\IEEEauthorblockA{\textit{Institute of Information Technology} \\
\textit{University of Dhaka}\\
Dhaka, Bangladesh \\
\texttt{bsse1104@iit.du.ac.bd}}
\and
\IEEEauthorblockN{Shartaz Sajid Nahid}
\IEEEauthorblockA{\textit{Institute of Information Technology} \\
\textit{University of Dhaka}\\
Dhaka, Bangladesh \\
\texttt{bsse1123@iit.du.ac.bd}}
\and
\IEEEauthorblockN{Md Arif Hasan}
\IEEEauthorblockA{\textit{Institute of Information Technology} \\
\textit{University of Dhaka}\\
Dhaka, Bangladesh \\
\texttt{bsse1112@iit.du.ac.bd}}
\linebreakand
\IEEEauthorblockN{Nurul Ahad Tawhid}
\IEEEauthorblockA{\textit{Institute of Information Technology} \\
\textit{University of Dhaka}\\
Dhaka, Bangladesh \\
\texttt{tawhid@iit.du.ac.bd}}
\and
\IEEEauthorblockN{Kazi Sakib}
\IEEEauthorblockA{\textit{Institute of Information Technology} \\
\textit{University of Dhaka}\\
Dhaka, Bangladesh \\
\texttt{sakib@iit.du.ac.bd}}
}

\maketitle

\begin{abstract}
Black-box Test Suite Minimization (TSM) techniques reduce testing costs without requiring access to production code. However, existing effective approaches rely on evolutionary search algorithms, introducing non-determinism that produces inconsistent results across runs, undermining reliability in automated testing pipelines. We propose DTM (Deterministic approaches for black-box Test suite Minimization), a framework that ensures deterministic test suite reduction while preserving effectiveness and efficiency. DTM converts test cases into Abstract Syntax Trees and computes pairwise similarities using four tree-based measures. For subset selection, it employs three deterministic algorithms: Modified Minimum Spanning Tree, Spectral Clustering, and Dynamic Programming. We evaluated DTM on 16 Java projects from Defects4J with 661 buggy versions. Experimental results show that DTM achieved an average accuracy of 0.74 with an execution time of just 0.98 minutes, outperforming all state-of-the-art approaches. Moreover, it consistently produced identical results across multiple runs, ensuring full determinism.

\end{abstract}


\section{Introduction}

Test Suite Minimization (TSM) aims to reduce the number of test cases while preserving their ability to detect faults \cite{DBLP:journals/stvr/YooH12, DBLP:journals/access/KhanLJA18}. TSM, performed without access to the production code, relying solely on test artifacts, is known as black-box. Black-box TSM techniques commonly use metrics such as test runtime, historical test logs, or test code similarity to estimate a test case’s fault detection potential. Recent work has explored Abstract Syntax Tree (AST)-based\cite{DBLP:conf/icse/PanGB23} and LLM-based\cite{10697930} similarity measures, showing promising effectiveness and practical efficiency. However, to select optimal subsets, these techniques often rely on evolutionary search strategies such as Genetic Algorithms (GA) or Non-dominated Sorting Genetic Algorithm (NSGA-II), which introduces a fundamental drawback, non-determinism\cite{HASANCEBI20101033}.

Non-deterministic algorithms can produce different outputs when run multiple times on the same input, leading to inconsistent test selection\cite{HASANCEBI20101033}. This unpredictability undermines the reliability of TSM, especially in automated testing pipelines where repeatability is essential. Moreover, it complicates debugging, benchmarking, and trust in the testing process\cite{bell2018deflaker, faria2017non}. In contrast, deterministic approaches consistently yield the same reduced suite given the same input, ensuring reproducibility and stability\cite{wildman2005dealing}. For a minimization approach to be suitable for black-box TSM, it must achieve determinism while still preserving high fault detection effectiveness and remaining efficient in terms of execution time.

Early black-box TSM techniques such as FastLane relied on historical data and metadata, limiting applicability in third-party or outsourced environments \cite{DBLP:conf/icse/PhilipBKMN19}. TF-IDF-based clustering approaches eliminated metadata dependence but struggled with structural granularity and language generalization \cite{DBLP:conf/icse/CrucianiMVB19}. To improve this, Pan et al. introduced similarity-based techniques that rely on AST structures \cite{DBLP:conf/icse/PanGB23} and large language models \cite{10697930}. These approaches demonstrated good accuracy and scalability. However, the use of heuristic search renders them non-deterministic. This variability limits their practical adoption in deterministic CI/CD workflows.

Several black-box TSM techniques have been proposed in the literature. Philip et al.\cite{DBLP:conf/icse/PhilipBKMN19} introduced the first approach using historical test logs, commit complexity, and change history to guide test reduction. To reduce dependency on test logs, Siam et al.\cite{siam2023exploratory} proposed change-proneness as a lightweight metric, relying only on version control data. Cruciani et al.\cite{cruciani2019scalable} applied clustering by converting test cases into TF-IDF vectors. To overcome structural limitations of such textual approaches, recent methods have explored AST-based and LLM-based similarity measures \cite{DBLP:conf/icse/PanGB23, 10697930}. While these approaches demonstrate strong accuracy and scalability, their reliance on heuristic search introduces non-determinism, resulting in inconsistent outcomes across runs.

To address these limitations, we propose a deterministic black-box TSM framework called \textit{DTM}. It preprocesses the test cases and converts each into an AST to preserve structural semantics. Then, it computes all pairwise similarities using four tree-based similarity measures - \textit{Top-down, Bottom-up, Combined (Top-down and Bottom-up)} and \textit{Tree Edit Distance}. To minimize the test suite, DTM selects the subset that yields the minimum total similarity using three known deterministic algorithms: modified Minimum Spanning Tree (MST), Spectral Clustering (SC), and Dynamic Programming (DP).

We evaluated DTM on 16 real-world Java projects containing 661 buggy versions. Experiments were conducted across 12 configurations and three minimization budgets (25\%, 50\%, and 75\%). Our best-performing configuration, SC with tree edit distance, achieved a 7\% accuracy gain and 50x times faster execution than the baseline ATM, and 2.9\% higher accuracy and 2.8 times speedup over LTM. More importantly, all DTM configurations produced identical results across multiple runs, unlike ATM and LTM, which varied due to non-determinism. 
These results highlight DTM’s potential as a reliable, effective, and efficient solution for black-box TSM.

\section{Related Work}

TSM aims to reduce the number of test cases while preserving the test suite’s ability to detect faults effectively \cite{DBLP:journals/access/KhanLJA18, DBLP:journals/jksucis/RhmannPAP20}. Depending on the type of information utilized, TSM techniques are broadly categorized as (1) white-box, which depend on production code, requirements, or models; and (2) black-box, which rely solely on test code or version control metadata \cite{DBLP:conf/icse/PanGB23}. The latter category, due to its lightweight nature, has gained popularity in industrial applications where access to source code is limited \cite{DBLP:conf/icse/PanGB23, 10697930, siam2023exploratory, DBLP:conf/icse/PhilipBKMN19}.

Philip et al.\ \cite{DBLP:conf/icse/PhilipBKMN19} introduced the first black-box TSM approach, FastLane, which used features such as commit complexity, test runtime, and historical logs to train a logistic regression model that prunes redundant test cases. Their approach achieved a 99.99\% fault retention with an 18.04\% reduction in test suite size. However, reliance on historical test logs restricts its applicability in setups where such data is unavailable \cite{DBLP:conf/icse/PanGB23}. To address this, Siam et al.\ \cite{siam2023exploratory} proposed a more lightweight solution that uses only commit logs to estimate the change-proneness of classes, scoring test cases based on their association with frequently changed components. While efficient, this method depends on version control metadata, making it unsuitable for outsourced or third-party testing environments where such access is not guaranteed.

Cruciani et al.\ \cite{cruciani2019scalable} attempted to eliminate metadata dependency altogether by using only the test code. They transformed test cases into TF-IDF vectors \cite{johnson1984extensions}, applied random projection\cite{turney2010frequency} for dimensionality reduction, and used clustering to select a diverse subset. This approach was efficient and effective for C projects, but it failed to generalize well to Java programs.

To overcome this limitation, Pan et al. proposed similarity-based techniques designed for Java. Their approach included two variants: one based on AST similarity\cite{DBLP:conf/icse/PanGB23} and another leveraging large language models (LLMs)\cite{10697930}. In the AST-based variant (ATM), each test case was parsed into an AST, and pairwise similarity was computed using tree-edit distance measures. In the LLM-based variant (LTM), each test case was embedded into a vector space, and pairwise similarity was calculated using cosine and Euclidean distances. Both variants used evolutionary search (GA and NSGA-II) to select a subset of test cases that minimized overall similarity. Although these techniques improved effectiveness and demonstrated scalability, their reliance on evolutionary search algorithms makes both the approaches non-deterministic\cite{HASANCEBI20101033}. This non-determinism introduces variability in the selected test subsets across runs, making the outcomes unreliable for practical and repeatable use\cite{faria2017non, wildman2005dealing}. Therefore, deterministic optimal subset selection algorithms need to be explored that maintain high effectiveness and efficiency while ensuring determinism.

\section{Methodology}\label{sec:methodology}
\vspace{-0.05cm}

In this study, we propose a black-box framework named DTM for TSM. The goal is to reduce test suites in a deterministic manner by leveraging tree-based similarity of test code, ensuring both effectiveness and efficiency. An overview of the core steps in the DTM framework is illustrated in Figure~\ref{fig:method}.
\begin{figure*}[h]
    \centering
    \includegraphics[width=1.9\columnwidth]{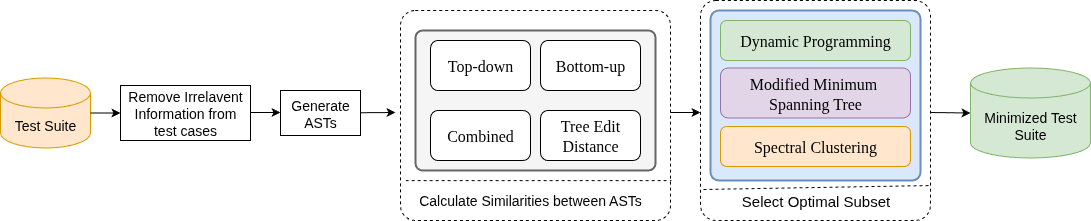}
    \caption{Overall framework of DTM}
    \label{fig:method}
    \vspace{-0.4cm}
\end{figure*}
\subsection{Remove irrelevant information from test cases}
To prepare test cases for similarity-based analysis, we first preprocess their source code of test cases by removing elements considered irrelevant to test behavior. Specifically, we remove test case names, Javadoc, single and multi-line comments, logging or printing statements, and test oracles such as assertions that oracles verify outcomes rather than drive execution, following Silva et al. \cite{DBLP:conf/sbcars/SilvaV20}. Additionally, we normalize variable identifiers while preserving data types by assigning sequential IDs (e.g., id1, id2) to maintain data flow and logic. This preprocessing ensures a cleaner and more relevant comparison of test cases. 

\subsection{Generate AST}

Processing test case code as natural language using text-based or token-based techniques fails to capture its syntactical information, leading to less accurate similarity measurements \cite{DBLP:conf/icse/ZhangWZ0WL19}. To address this, we use ASTs to preserve the syntactic structure of test case code. Utilizing an AST parser from the Eclipse JDT\footnote{\url{https://projects.eclipse.org/projects/eclipse.jdt}} library, we statically traverse and transform the test case code into corresponding ASTs \cite{DBLP:journals/cl/Noonan85}. Comparing these ASTs allows us to identify precise differences, such as variations in method calls, parameters, and their values. 

\subsection{Calculate similarities between ASTs}
\label{sec: similiarity}

A single similarity measure may perform inconsistently across projects, as different measures capture distinct structural information of the test cases. Therefore, following Valiente et al. \cite{DBLP:series/txcs/Valiente21}, we adopt four complementary methods: (i) Top-down similarity, which prioritizes high-level structure of test cases first based on the top-down maximum ordered common subtree isomorphis, (ii) Bottom-up similarity, which emphasizes low-level details like method call parameters, (iii) Combined measure, that merges both top-down and bottom-up measures, and (iv) Tree Edit Distance, which captures scattered structural changes by computing the minimal set of edits needed to transform one tree into another. These methods provide robust similarity measurements suited to a range of test case structures. The detailed formulations of these similarity measures can be found in the work of Valiente et al. \cite{DBLP:series/txcs/Valiente21}.


\subsection{Selecting an optimal subset of test cases}

Now that we have captured the similarities between all test cases, we want to select a subset of size $k$ from the $n$ test cases based on their similarities, where $k<n$. Specifically, our objective is to minimize the total pairwise similarity within the subset to maximize its diversity and reduce redundancy. We formulate this as an optimal subset selection problem and explore three distinct approaches: \emph{Modified Minimum Spanning Tree (MST)}, \emph{Spectral Clustering (SC)}, and \emph{Dynamic Programming (DP)}. These methods leverage different techniques such as graph theory and linear algebra, to identify a representative and diverse subset of test cases.

\subsubsection{Modified Minimum Spanning Tree (MST)}




To obtain a reduced test-suite with minimal redundancy, we adopt the MST algorithm (Prims), similar to \cite{tarek2022clustering}, and modify it. First, we select the least similar pair of all test cases and add them to the selected subset. Then, we iteratively add a new test case with minimum \emph{loss value} until the size of the selected subset is $k$. Here, we define the \emph{loss value} of a node as its maximum similarity score with any test case in the currently selected subset. This selection process ensures both diversity and relevance among the chosen test cases. This, in turn, enhances testing efficiency and facilitates more effective software testing practices across iterative development cycles. 

To illustrate the algorithm, consider a similarity matrix $S$ where each entry $s_{ij}$ represents the similarity score between test cases $i$ and $j$, as shown in Figure~\ref{fig:example}. The algorithm begins by identifying the pair of test cases with the minimum similarity score. In this example, test cases 1 and 3 exhibit the lowest similarity score of 0.1 among all pairs, and are therefore selected as the initial subset. Next, we calculate the loss value for each test case in the unselected subset. For instance, test case 2 has similarity scores with selected cases 1 and 3 as [0.5, 0.4]. The loss value for test case 2 would be 0.5. Similarly, loss value of test case 4 and 5 would be 0.5 and 0.6, respectively. Among the unselected test cases (2, 4, 5), test case 4 has the minimum loss value of 0.4. Therefore, we insert test case 4 into the selected subset. To find the whole subset, we repeat this process until $k$ test cases are selected. 
    
\subsubsection{Spectral Clustering}



SC \cite{tutorial} offers an alternative approach for our subset selection problem. SC is particularly designed to group data with strong pairwise similarities into same clusters by analyzing a similarity graph. This makes it ideal for selecting diverse and representative test cases based on their pairwise similarities, as we already have a representative similarity matrix among the ASTs. To start the process, we first compute the Laplacian matrix \( L = D - S \), where \( S \) is the similarity matrix and \( D \) is the degree matrix. Since our objective is to select an optimal subset of $k$ diverse test cases, we take eigenvectors of \( L \) corresponding to the \( k \) smallest eigenvalues. Based on these selected $k$ eigenvectors, we divide all test cases into $k$ clusters such that intra-cluster similarity is maximized and inter-cluster similarity minimized.

Now that we have $k$ diverse clusters, we construct the final subset by selecting one representative test case from each of the $k$ clusters. Since test cases within the same cluster are highly similar, this process ensures that the selected subset is diverse and minimally redundant. By leveraging the spectral properties of the similarity graph, this approach naturally balances coverage and diversity.



\begin{figure}[h]
        \centering
        \includegraphics[width=0.6\columnwidth]{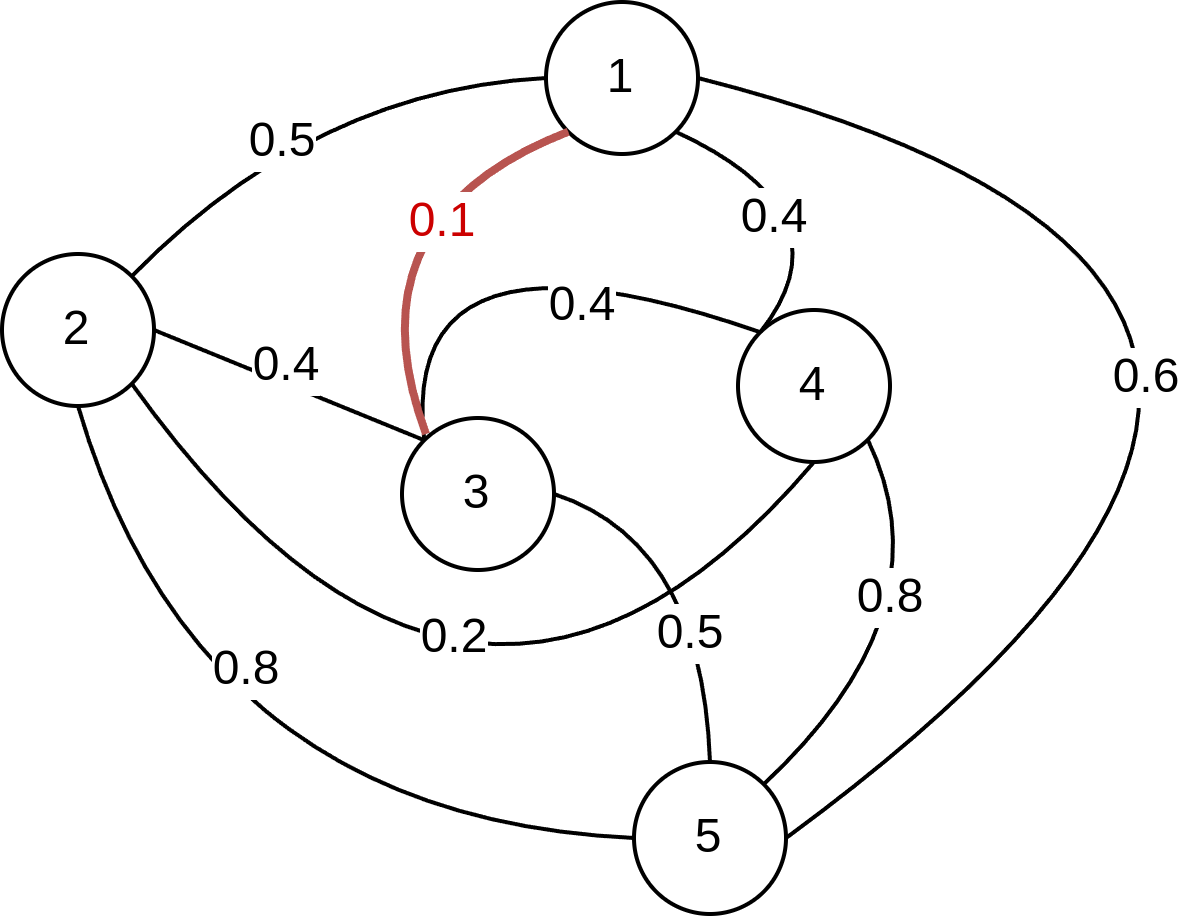}
        \caption{Example of Modified MST}
        \label{fig:example}
        \vspace{-0.35cm}
\end{figure}

\subsubsection{Dynamic Programming}

This approach uses dynamic programming to select the desired subset of size $k$ in a way that the total pairwise similarity within the selected subset is minimized. To achieve this, we construct a DP table, where each entry $\text{dp}[i][j]$ represents the minimum total similarity when selecting $j$ test cases from the first $i$ test cases, where $j\le k$ and $i\le n$. At each step, we check whether to include or exclude the current test case based on the resulting similarity, which is computed between the current test case and the already selected subset. In this manner, it stores the best possible solution for smaller sub-problems. After filling the DP table, the selected subset is constructed by tracing back the decisions from the DP table. This structured method aims to produce a subset with minimal redundancy.






\section{Validation}
In this section, we evaluate our proposed approaches through a series of experiments. We begin by outlining the experimental setup, then we describe the systems under test and the evaluation metrics. Finally, we analyze the results obtained from applying our approaches to the test suites.

\begin{table*}[htbp]
  \centering
  \caption{Accuracy Statistics across all Approaches under 50\% Minimization Budget}
  \label{tab:results}
  \resizebox{\textwidth}{!}{%
  \begin{tabular}{|l|c|c|c|c|c|c|c|c|c|c|c|c|c|c|}
    \hline
    \multirow{2}{*}{\textbf{Statistic}} & \multicolumn{4}{c|}{\textbf{Dynamic Programming}} & \multicolumn{4}{c|}{\textbf{MST}} & \multicolumn{4}{c|}{\textbf{Spectral Clustering}} & \textbf{ATM} & \textbf{LTM} \\
    \cline{2-15}
    & \begin{tabular}[c]{@{}c@{}}\textbf{Top}\\\textbf{Down}\end{tabular} & 
      \begin{tabular}[c]{@{}c@{}}\textbf{Bottom}\\\textbf{Up}\end{tabular} & 
      \begin{tabular}[c]{@{}c@{}}\textbf{Comb}\\\textbf{ined}\end{tabular} &
      \begin{tabular}[c]{@{}c@{}}\textbf{Tree}\\\textbf{Edit}\end{tabular} & 
      \begin{tabular}[c]{@{}c@{}}\textbf{Top}\\\textbf{Down}\end{tabular} & 
      \begin{tabular}[c]{@{}c@{}}\textbf{Bottom}\\\textbf{Up}\end{tabular} & 
      \begin{tabular}[c]{@{}c@{}}\textbf{Comb}\\\textbf{ined}\end{tabular} &
      \begin{tabular}[c]{@{}c@{}}\textbf{Tree}\\\textbf{Edit}\end{tabular} &
      \begin{tabular}[c]{@{}c@{}}\textbf{Top}\\\textbf{Down}\end{tabular} & 
      \begin{tabular}[c]{@{}c@{}}\textbf{Bottom}\\\textbf{Up}\end{tabular} & 
      \begin{tabular}[c]{@{}c@{}}\textbf{Comb}\\\textbf{ined}\end{tabular} &
      \begin{tabular}[c]{@{}c@{}}\textbf{Tree}\\\textbf{Edit}\end{tabular} & 
      \begin{tabular}[c]{@{}c@{}}\textbf{Tree}\\\textbf{Edit/GA}\end{tabular} &
      \begin{tabular}[c]{@{}c@{}}\textbf{UniXcoder/}\\\textbf{Cosine Sim}\end{tabular} \\
    \hline
    \textbf{Min} & 0.25 & 0.45 & 0.25 & 0.25 & 0.52 & 0.36 & 0.46 & 0.36 & 0.48 & 0.45 & 0.52 & \textbf{0.59} & 0.37 & 0.53 \\
    \hline
    \textbf{25 Percentile} & 0.45 & 0.48 & 0.45 & 0.48 & 0.60 & 0.54 & 0.55 & 0.57 & 0.59 & 0.59 & 0.61 & 0.64 & 0.62 & \textbf{0.65} \\
    \hline
    \textbf{Mean} & 0.57 & 0.63 & 0.57 & 0.56 & 0.68 & 0.64 & 0.64 & 0.63 & 0.66 & 0.69 & 0.68 & \textbf{0.73} & 0.67 & 0.71 \\
    \hline
    \textbf{Median} & 0.56 & 0.60 & 0.56 & 0.52 & 0.66 & 0.62 & 0.65 & 0.62 & 0.66 & 0.64 & \textbf{0.72} & \textbf{0.72} & 0.66 & 0.70 \\
    \hline
    \textbf{75 Percentile} & 0.66 & 0.75 & 0.66 & 0.61 & 0.75 & 0.74 & 0.73 & 0.75 & 0.76 & \textbf{0.82} & 0.76 & 0.81 & 0.74 & 0.73 \\
    \hline
    \textbf{Max} & \textbf{1.00} & 0.75 & 0.75 & 0.75 & \textbf{1.00} & \textbf{1.00} & 0.77 & 0.81 & 0.81 & \textbf{1.00} & 0.82 & \textbf{1.00} & 0.91 & \textbf{0.84} \\
    \hline
  \end{tabular}
  }
\end{table*}

\begin{table}[htbp]
  \centering
  \caption{Execution Time under 50\% Minimization Budget}
  \label{tab:time-results}
  \begin{tabular}{|l|c|c|c|c|c|}
    \hline
    \textbf{Statistic} & \textbf{DP} & \textbf{MST} & \textbf{SC} & \textbf{ATM} & \textbf{LTM} \\
    \hline
    \textbf{Min} & \textbf{0.01} & \textbf{0.01} & \textbf{0.01} & 0.39 & 0.16 \\
    \hline
    \textbf{25 Percentile} & 0.26 & \textbf{0.03} & 0.05 & 1.34 & 0.38 \\
    \hline
    \textbf{Mean} & 3.46 & 3.21 & \textbf{0.92} & 45.80 & 2.60 \\
    \hline
    \textbf{Median} & 0.71 & 0.31 & \textbf{0.23} & 4.12 & 0.53 \\
    \hline
    \textbf{75 Percentile} & 2.37 & 3.29 & \textbf{1.06} & 43.65 & 2.93 \\
    \hline
    \textbf{Max} & 27.66 & 32.62 & \textbf{7.28} & 320.40 & 12.75 \\
    \hline
  \end{tabular}
\end{table}

\subsection{Experimental Design and Dataset}


We evaluated the performance of our proposed approaches across 12 configurations, combining three deterministic algorithms with four test similarity measures. Experiments were run on an Intel Core i5-9400F (6 cores, 2.9GHz), 16GB RAM, Ubuntu 22.04. Following prior work \cite{DBLP:conf/icse/PanGB23, siam2023exploratory, 10697930}, we used the Defects4J dataset\footnote{https://github.com/rjust/defects4j}, which includes 661 buggy versions across 16 real-world Java projects. The project sizes range from 2 KLoC to 74 KLoC, with test suites spanning 4 KLoC to 73 KLoC. Each project has 4 to 174 buggy versions, with 152 to 3,919 test cases per version on average. We used the published replication packages of ATM and LTM \footnote{\url{https://zenodo.org/records/13685828},\url{https://zenodo.org/records/7455766}} to reproduce their results and compare them with ours. To reduce result variability of the observed results, we run each configuration 10 times and report the average findings.

\subsection{Minimization Budgets} \label{min_budget}

The minimization budget specifies the target size of the reduced test suite as a percentage of the original. Following prior work \cite{DBLP:conf/icse/PanGB23,10697930}, we used 25\%, 50\%, and 75\% budgets, reflecting common industry practices \cite{DBLP:conf/icse/PanGB23}.

\subsection{Evaluation Metrics}
To measure effectiveness, we used \emph{accuracy}, which reflects the proportion of faults preserved in the reduced suite. It is calculated as shown in Eq. \ref{eq:accuracy}:
\vspace{-0.1cm}
\[
\text{Accuracy} = \frac{|F'|}{|F|} \tag{3}\label{eq:accuracy}
\]
\noindent where \(|F'|\) represents the number of fault-revealing tests in the reduced suite and \(|F|\) is the total number of faults in the original test suite. For efficiency, we used \emph{execution time}.

\subsection{Results and Discussion}

We obtained results using our methodology for all three minimization budgets stated in \ref{min_budget}. However, in this section, we focus on the results achieved under the 50\% minimization budget similar to \cite{DBLP:conf/icse/PanGB23, siam2023exploratory, 10697930} as results for the other minimization budgets were consistent. The data, subject project information, codes and results obtained can be found in our replication package\footnote{\url{https://figshare.com/s/853d3729a5d70b15a63e}}.

\begin{figure}
    \centering
    \includegraphics[width=1\linewidth]{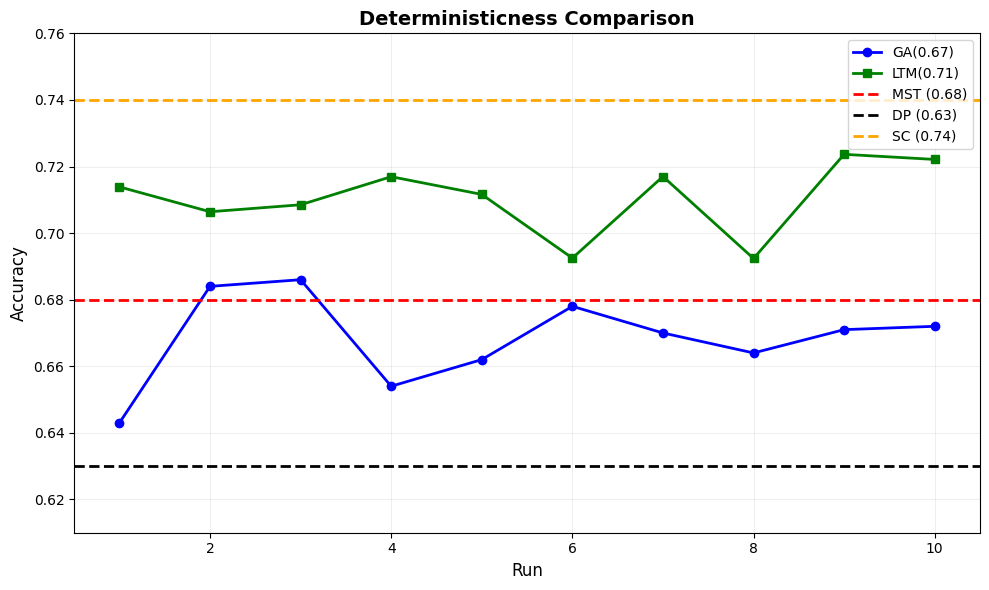}
    \caption{Avg accuracy of the approaches over different runs}
    \label{fig:deterministic}
    \vspace{-0.7cm}
\end{figure}

\textbf{\textit{Effectiveness.}} Table~\ref{tab:results} demonstrates accuracy statistics across three optimization algorithms, DP, MST, and SC, using four similarity measures, alongside baselines ATM and LTM for a 50\% minimization budget. Bold values mark the best-performing setups. The results show that similarity measures significantly affect algorithm effectiveness, with tree edit distance consistently delivering the highest accuracy.

Among the algorithms, SC with tree edit distance achieved the strongest performance, with a mean accuracy of 0.73 and median of 0.72. This reflects SC’s ability to model complex, non-linear test dependencies combined with the precise structural comparison of tree edit distance. In contrast, DP performed weakest, with mean accuracy between 0.56 and 0.63. Its poor result on project \texttt{Collections} (accuracy 0.25) highlights DP’s limitation in handling complex inheritance of projects, due to its greedy nature. Interestingly, the bottom-up similarity within DP showed relatively better stability (mean 0.63), suggesting that building similarity from smaller components upward offers more consistent guidance than top-down methods. MST showed balanced but variable performance, with tree edit distance again best (mean 0.63). However, for project \texttt{Time}, MST failed completely with bottom-up and combined similarities (accuracy 0.00) but achieved perfect accuracy (1.00) with tree edit distance. This illustrates MST’s sensitivity to similarity quality and the robustness of structural measures in guiding spanning tree construction.

Configuration SC combined with tree edit distance outperformed both baselines, achieving 2.9\% and 7\% accuracy gain over LTM (0.71) and ATM (0.67), respectively. This demonstrates that combining structural similarity with sophisticated clustering provides better guidance than heuristic search–based optimization in TSM.

\textbf{\textit{Efficiency.}} Table~\ref{tab:time-results} reports the execution times across different approaches. SC achieved the highest overall efficiency, with a mean execution time of 0.92 minutes and a median of 0.23 minutes, outperforming both baseline techniques. Even for the largest project, \textit{Time}, which includes 3918 test cases, SC required only 7.28 minutes, highlighting its suitability for large-scale industrial applications. Both DP and MST demonstrated similar average performance, with mean times of 3.46 and 3.21 minutes, respectively. However, MST showed higher variance, ranging from 0.23 to 32.62 minutes across projects. Overall, SC's efficiency translates to a 50 times speedup over ATM and a 2.82 times improvement over LTM.

\textbf{\textit{Determinism.}} As demonstrated in Figure~\ref{fig:deterministic}, over 10 independent runs, heuristic search-dependent algorithms ATM and LTM showed variable accuracy results, with ATM ranging from 0.643 to 0.686 (standard deviation: 0.015) and LTM ranging from 0.692 to 0.724 (standard deviation: 0.012). In contrast, all three graph-based approaches (DP, MST, and SC) exhibited deterministic behavior, producing identical minimized test suites and therefore consistent accuracy across all runs. This deterministic property ensures reproducible results and eliminates the uncertainty inherent in stochastic optimization methods, making our approaches more suitable for automated testing pipelines where consistency is critical.

\begin{center}
\fbox{%
  \parbox{\dimexpr1.0\linewidth-2\fboxsep-2\fboxrule}{%
    \textbf{\textit{Summary of Findings.}}
    All our proposed approaches show deterministic behavior while maintaining strong efficiency and effectiveness. In particular, Spectral Clustering with tree edit distance outperformed both baselines, ATM and LTM, with up to 7\% higher accuracy and 50 times faster execution. These results demonstrate that our method provides reliable, effective, and efficient test suite minimization.
  }%
}
\end{center}

\section{Threats to Validity}

Our evaluation is based solely on the Defects4J dataset, which, while widely adopted in TSM research, may limit external validity. The results may not fully generalize to projects with different structures, domains, or testing practices. Furthermore, since all test cases are written in Java, our findings may not extend to test suites in other programming languages, potentially affecting language generalizability. Although all experiments were conducted on the same hardware and software configuration and each was repeated 10 times to account for variability, non-deterministic techniques such as ATM and LTM may still yield minor result fluctuations, introducing a threat to internal validity.


\section{Conclusion and Future Work}

In this study, we presented DTM, a deterministic framework for black-box test suite minimization based on tree-structured test code similarity. Unlike prior approaches that rely on heuristic search, DTM ensures consistent outputs by applying three deterministic algorithms, MST, SC, and DP, to select a subset of test cases with minimal total similarity. Experimental results across 16 real-world projects containing 661 buggy versions demonstrate that DTM not only achieves deterministic behavior but also improves efficiency and maintains high fault detection effectiveness. These findings highlight DTM's practical suitability for reliable and reproducible test automation workflows. 

Future work can investigate the impact of additional deterministic strategies, such as optimization-based and hybrid approaches. Additionally, the similarity matrix can be treated as a tensor \cite{akhter2024low} to better capture the relationships among the test cases better. Moreover, assessing performance across various programming languages can help further validate the approach.



\small 
\bibliography{CP-TSM} 
\bibliographystyle{IEEEtran}

\end{document}